\documentclass[pss]{revtex4-1} 
\usepackage{amsmath}
\usepackage{bm}             
\usepackage{graphicx}

\begin{document}

\title{Electronic and Optical Properties of Functionalized GaN (10-10) Surfaces using Hybrid-Density Functionals}

\author{Dennis Franke}
\affiliation{Bremen Center for Computational Materials Science, University of Bremen, Am Fallturm 1, 28359 Bremen, Germany}
\author{Andreia Luisa da Rosa}
\affiliation{Universidade Federal de Goi\'as, Campus Samambaia, 74690-900 Goi\^ania, Brazil}
\affiliation{Bremen Center for Computational Materials Science, University of Bremen, Am Fallturm 1, 28359 Bremen, Germany}
\author{Michael Lorke}
\affiliation{Bremen Center for Computational Materials Science, University of Bremen, Am Fallturm 1, 28359 Bremen, Germany}
\author{Thomas Frauenheim}
\affiliation{Bremen Center for Computational Materials Science, University of Bremen, Am Fallturm 1, 28359 Bremen, Germany}


\begin{abstract}Adsorption of small ligands on semiconductor surfaces is
  a possible route to modify these surfaces so that they can be used
  in biosensing and optoelectronic devices. In this work we perform
  density-functional theory calculations of electronic and optical
  properties of small ligands on GaN-(10$\bar1$0) surfaces. From the
  investigated anchor groups we show that thiol groups introduce
  states into the GaN band gap. However, these state are not optically
  active, at least for these perfect surfaces. This means that more
  realistic surfaces need to be considered to suggest how surface
  modification can enhance the optical properties of GaN non-polar
  surfaces.
  \end{abstract}

\maketitle   

\section{Introduction}

Scientific interest in hybrid nanostructures consisting of organic and
inorganic materials for the fabrication of electronic and
optoelectronic devices with novel properties has grown over the past
years\,\cite{STVR,STVR2}. GaN is a semiconductor material with a wide
(3.4 eV) band gap widely used in ultraviolet-blue light emitting
diodes, photodetectors and lasers \cite{Nakamura,Morkoc}. Due to its
stability one of the major interest in GaN is for biomedical
applications\cite{Guo,Kim2006,Williams,Choi2017}.  Functionalization
with small ligands is a possible route to functionalize semiconductor
surfaces, since these groups can subsequently bind covalently to a
wide range of biomolecules or even be used to introduce optically
active states in the band gap which are suitable for optoelectronic
devices. Several organic groups such as thiols \cite{Bermudez},
alkenes \cite{Ambacher} and silanes\cite{Lieberman} have been used to
modify the surface of GaN. In particular, the adsorption of thiols on
GaN surfaces indicates that the molecule adsorbs via the thiol group
and remain stable upon annealing\,\cite{Bermudez}. Furthermore, Stine
{\rm et al.}  \cite{Stine} developed a technique to produce amine
groups directly on GaN surfaces to successfuly imobilize biomolecules. However,
many aspects regarding suitable groups and their effect on the
electronic and optical properties of GaN remain unclear. In this work
we perform density-functional theory calculations of electronic and
optical properties of small ligands on GaN-(10$\bar1$0) surfaces as
they can be used to imobilize larger organic and bio-molecules or even
to modify the GaN surface to be used in optoelectronic devices. We
show that although thiol groups introduce states in the GaN band
gap. However, these state are not optically active. As we consider
only perfect surfaces, we suggest that more realistic surfaces need to
be considered to identify possible mechanisms to enhance the optical
properties of GaN non-polar surfaces.

\section{Computational Details}

 While structural properties such as bond lengths are reliably
 obtained within the PBE \,\cite{Perdew:96,Perdew:97} form of the
 exchange-correlation functional, the electronic structure can be
 severely underestimated\, \cite{GGAGap,GGAGap1}.  In particular for
 GaN, the use of PBE yields a band gap of 2.4 eV, a value which is too
 small compared to the experiment (3.2-3.4 eV)\,\cite{CRC}. It has
 been shown that the use of hybrid-density functional where an amount
 of Hartree-Fock exchange is added to the PBE exchange energy term
 lead to an overall improvement of chemical and physical properties of
 solids and molecules, specially band gaps\,\cite{Paier}. In this work
 we use the PBE0 form for the exchange correlation functional where
 25\% of Hartree-Fock exchange is added to the PBE functional \,\cite{PBE0} to
 investigate the structural and electronic properties of hybrid
 GaN-organic interfaces.  The ground state of GaN is a wurtzite
 structure with 4 atoms per unit cell. To model the surface, we
 consider a GaN slab containing 8 GaN layers with a vacuum region of
 18 {\AA}.  We adsorb two molecules per supercell, one on each side of
 the slab. Molecules of the form ${\rm CH_{3}-X}$ where X = COOH, SH and
 ${\rm NH_{2}}$ were adsorbed on the GaN surface with 1 ML (monolayer)
 coverage.  We have used density functional theory (DFT)
 \cite{Hohenberg1964,Kohn1965} as implemented in the Vienna ab initio
 simulation package (VASP) \cite{VASP:3,VASP:4}. The projected
 augmented wave (PAW) method has been employed \cite{Kresse:99,Bloechel:94}. The
 atomic structure was optimized with the PBE functional, while the electronic and optical properties have been calculated with the
 PBE0 functional. A plane wave basis with an energy cutoff
 of 400\,eV and a ($1 \times 10 \times 10$) Monkhorst-Pack {\bf
   k}-point sampling have been used in all calculations. Forces on all atoms hav been converged until they are smaller than 0.001 eV/{\AA}

\section{Results}

\begin{figure}[ht!]
  \centering
  \begin{tabular}{ccc}
\includegraphics[width=0.25\columnwidth,clip]{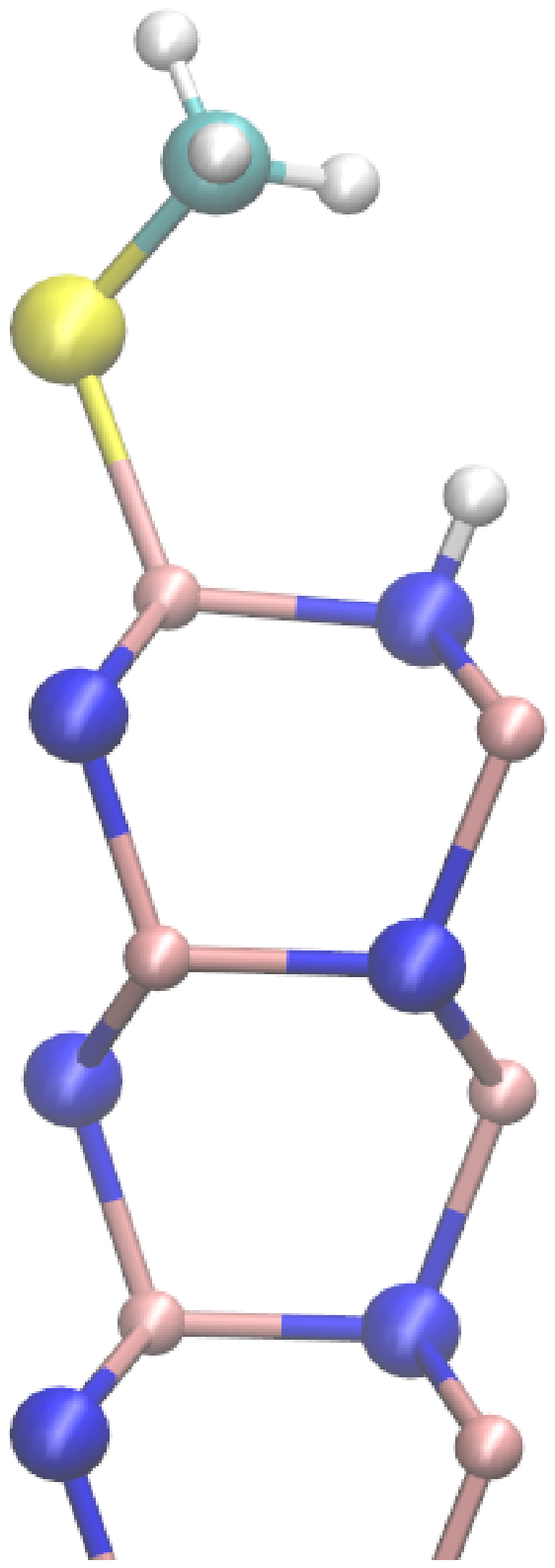}&
\includegraphics[width=0.31\columnwidth,clip]{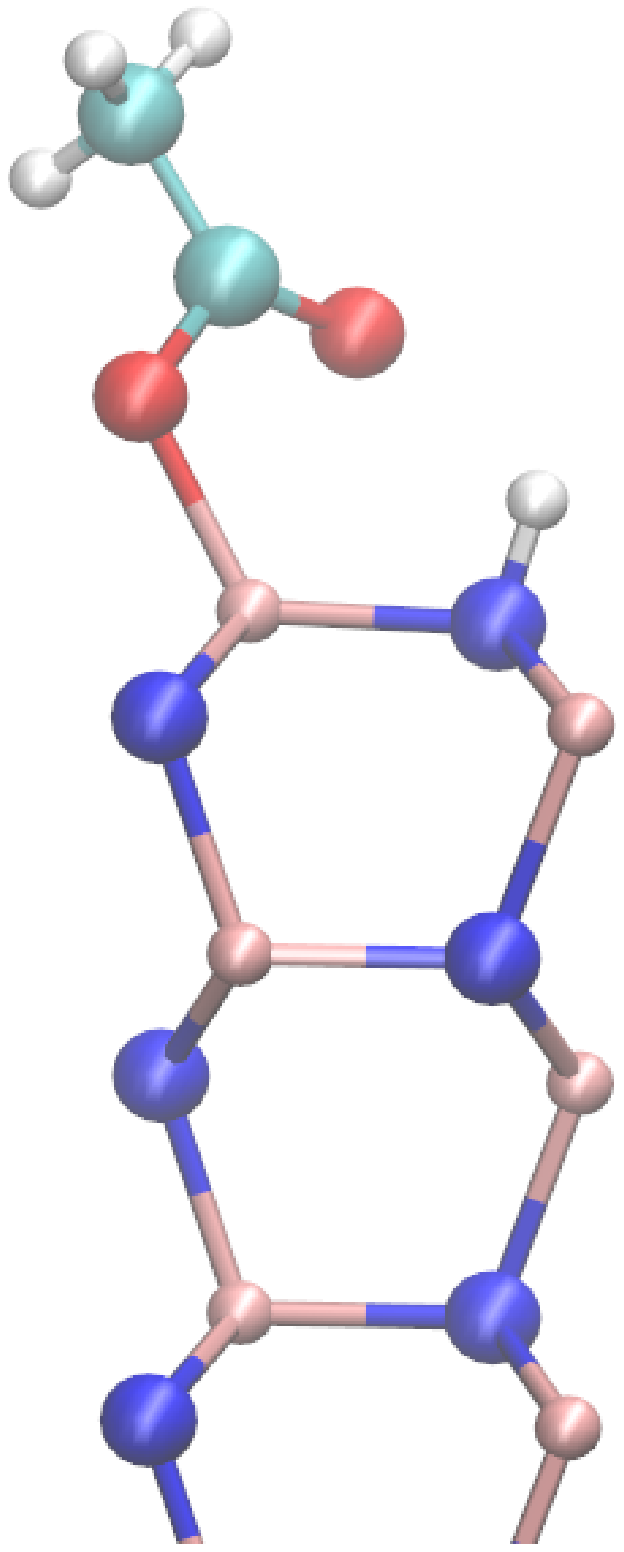}&
\includegraphics[width=0.34\columnwidth,clip]{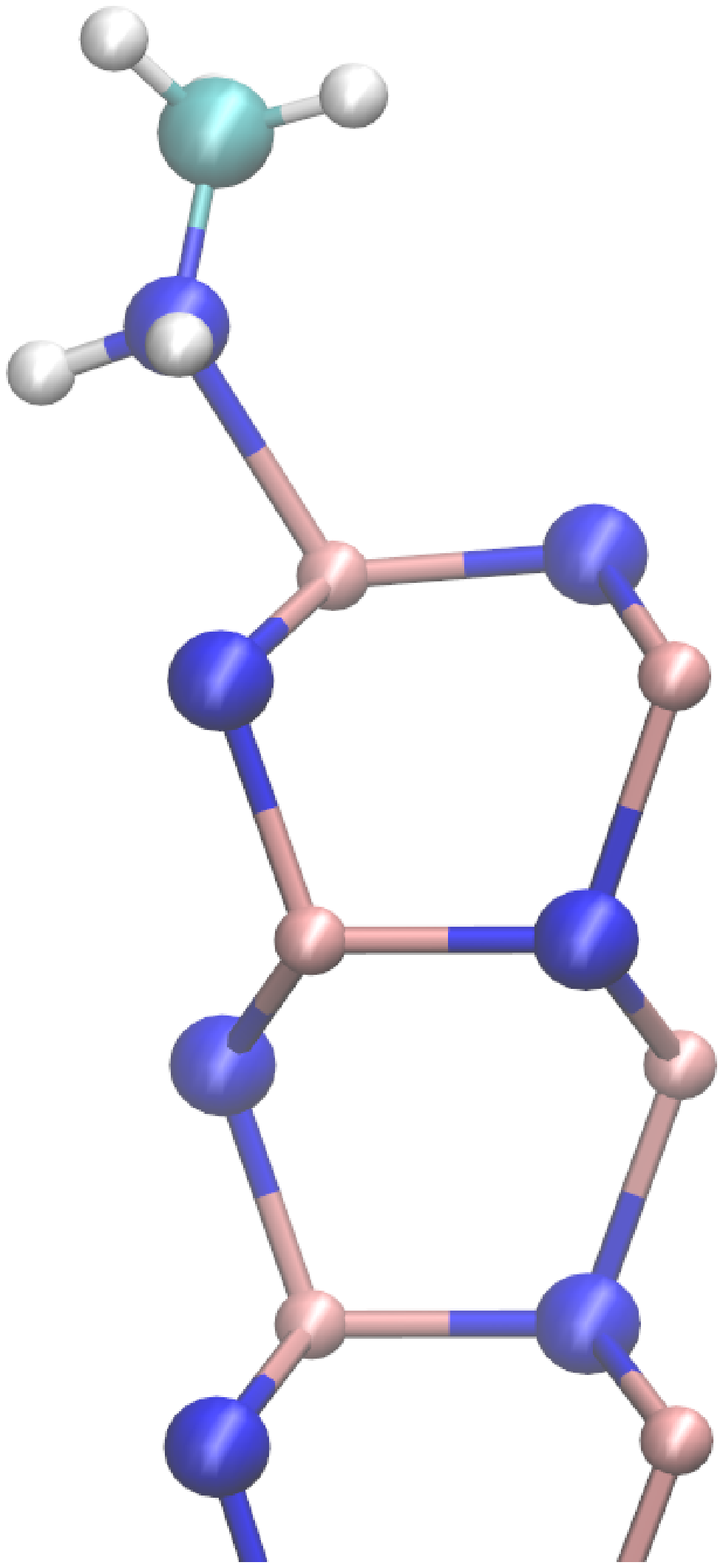}
\end{tabular}
\caption{Optimized structures of the GaN surface functionalized with ${\rm CH_{3}-SH}$ (left), ${\rm CH_{3}-COOH}$ (middle) and ${\rm CH_{3}-NH_{2}}$ (right) molecules.
The spheres represent gallium (light pink), nitrogen (blue), oxygen (red),  carbon (turquoise), sulfur (yellow) and hydrogen (white) atoms.}
\label{geometries}
\end{figure}

The optimized geometries of functionalized surfaces are shown in
Fig.~\ref{geometries}. Thiol and carboxyl groups adsorb on the surface in a dissociative
manner, while the amine group does not dissociate. All molecules
adsorb in a monodentate binding mode. The Ga-N bond lengths in the
center of the slab are similar to those found in bulk GaN, varying
between 1.95 and 2.00 {\AA}. This indicates that our slab is large
enough to avoid spurious interactions between the two sides of the
slab.  As the surface is modified with the small molecules, there is a
shortening of these bond lengths (1.84-1.90 {\AA}) for the $-{\rm NH_{2}}$ group
which binds non-dissociatively. On the other hand adorption of -SH
and -COOH groups lead to Ga-N bond lengths close to the bulk values.

For thiol groups (left figure) in Fig. ~\ref{geometries}, the -SH
groups adsorbs on the surface via S-Ga bonds with a bond length of
2.26 {\AA} and an angle of 115$^\circ$. The Ga-S-C angle is
111$^\circ$.  The carboxyl groups (Fig. ~\ref{geometries}, middle)
bind via two asymmetric bonds between the O atoms of the functional
group and the surface Ga atoms with a bond length of 1.90 {\AA} and a
binding angle of 120$^\circ$. The characteristic O-C-O angle is
122$^\circ$.  The amine groups adsorb via N-Ga bonds with a resulting
bond length of 2.15 {\AA} and an angle of 117$^\circ$. The Ga-N-C
angle is 132$^\circ$.  The discussed bond lengths are somewhat similar
to the results found in our previous work on functionalized ZnO
($10\bar{1}0$) surfaces \cite{Prev1,Prev2}. All groups bind strongly
to the GaN surfaces, which corroborates with previous experimental
findings which suggested that these groups are stable on
GaN\,\cite{Bermudez,Stine}.

\begin{figure}[ht!]
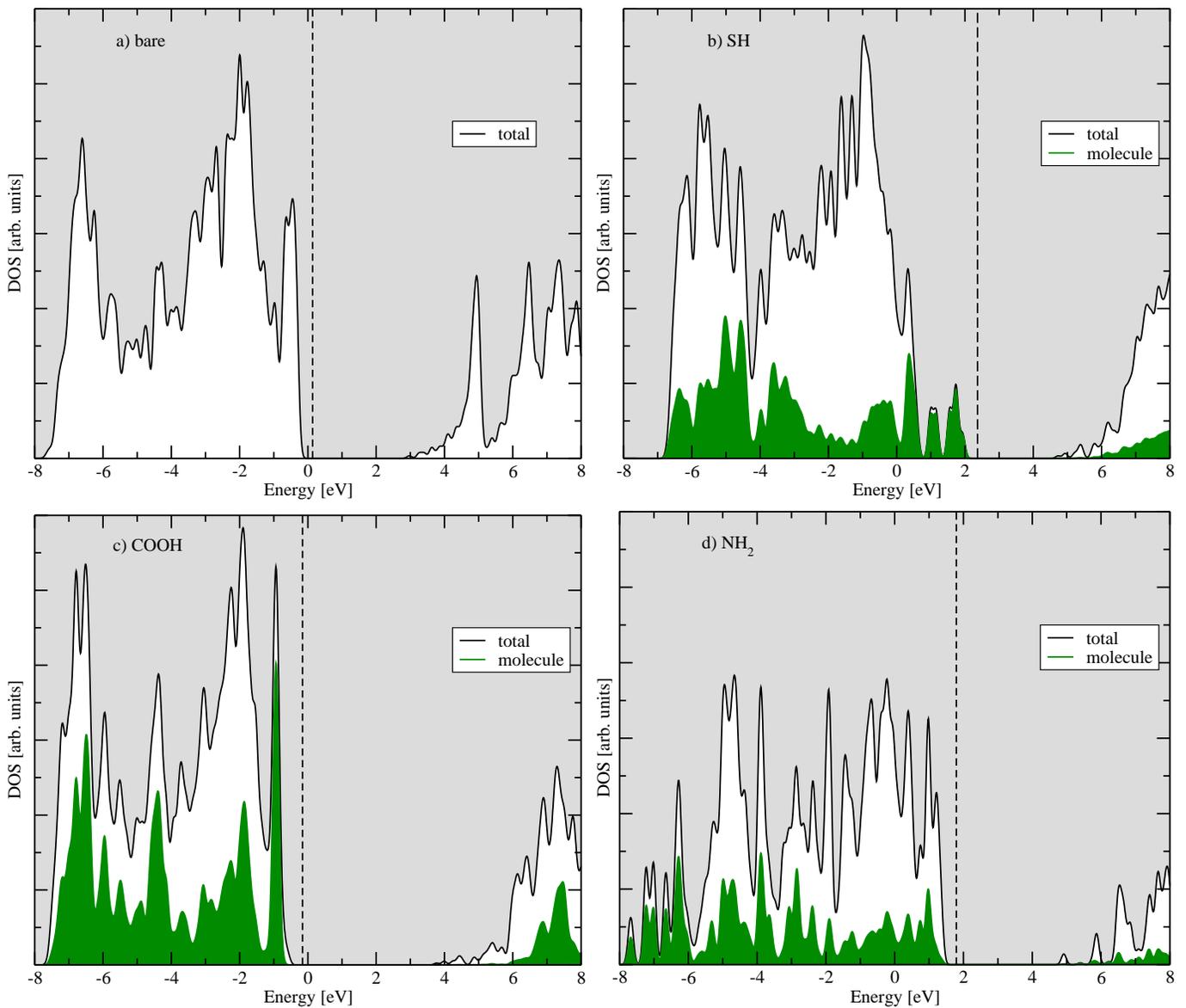

  \centering
  \begin{tabular}{cc}
\includegraphics[width=0.5\columnwidth,clip,keepaspectratio]{./DOS_BARE.eps}&
\includegraphics[width=0.5\columnwidth,clip,keepaspectratio]{./DOS_SH.eps}\\
\includegraphics[width=0.5\columnwidth,clip,keepaspectratio]{./DOS_COOH.eps}&
\includegraphics[width=0.5\columnwidth,clip,keepaspectratio]{./DOS_NH2.eps}
\end{tabular}
\caption{Projected DOS for the a) bare and functionalized surfaces with b) -SH, c) -COOH and d) -${\rm NH_{2}}$ groups. The full black line represents the total DOS, the green region shows the projection onto the  ligand states. The dashed lines denotes the Fermi energy.}
\label{Dos}
\end{figure}

\begin{figure}[]
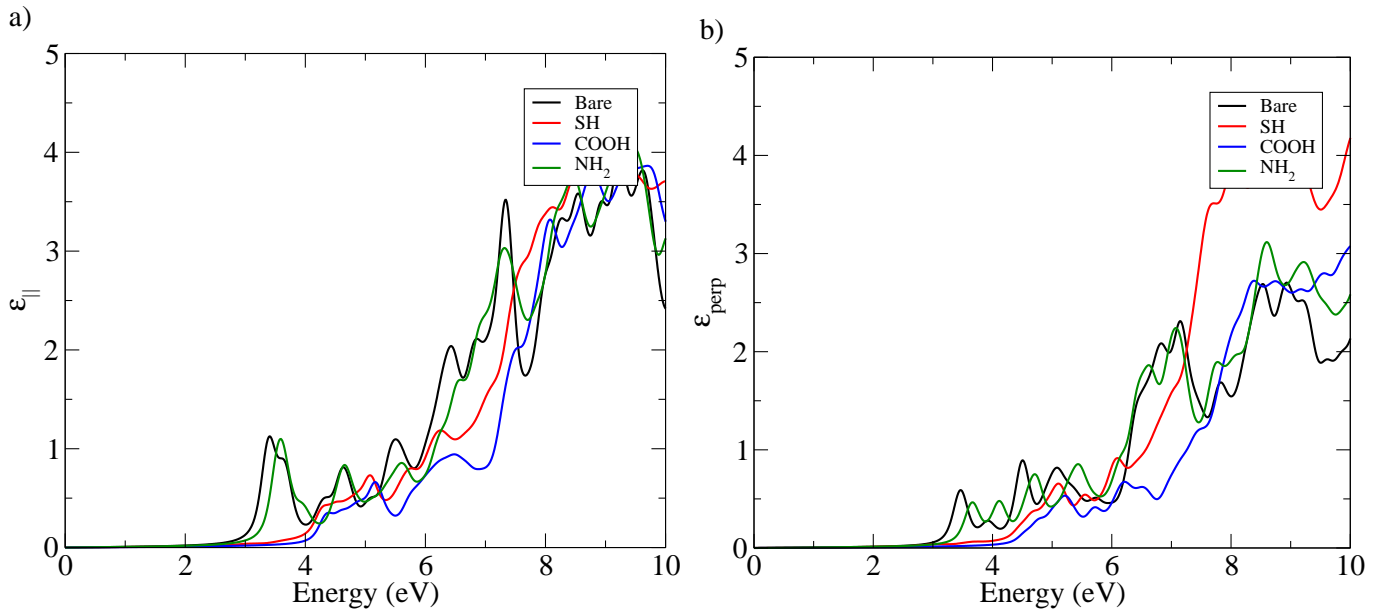

  \centering
  \begin{tabular}{cc}
\includegraphics[width=0.5\columnwidth,clip]{./DielParallel.eps} &
\includegraphics[width=0.5\columnwidth,clip]{./DielPerpendicular.eps}
\end{tabular}
\caption{Dielectric function for the bare and modified surfaces, 
shown are (a) $\varepsilon_{||}$ and (b) $\varepsilon_{\perp}$.}
\label{fig:diel}
\end{figure}

Next we discuss the electronic and optical properties of the
functionalized structures. The total density-of-states (DOS) and the
projected DOS onto the molecular orbitals for the bare and
functionalized surfaces are shown in Fig.\,\ref{Dos}. The band
alignment has been done with respect to the vacuum level as described
in Refs. \cite{Align1,Align2}. The use of the hybrid PBE0 functional
significantly improves the band gap of bulk GaN, yielding a value of
3.6 eV, somewhat overestimated but closer to experiment \cite{CRC} and
in agreement with previous calculations\cite{Scuseria}. We then
performed further electronic structure calculations for the bare and
modified surfaces with this functional. Bare surfaces have a slightly
reduced band gap compared to bulk value, due to extra surface states
appearing close to the valence and conduction bands (i have to check
this). Modified surfaces with -SH introduce intra-gap states, mainly
due to S-$p$ orbitals. These states are located 0.3\,eV and 1\,eV
above the top of the valence band (VBM), as it can be seen in
Fig.\,\ref{Dos} (b), similarly to what we have found for ZnO surfaces
\cite{Prev1}.  These state also hybridize with the N-$p$ surface
states, but this induces only small changes close to VBM.

On the other hand, functionalization with -COOH does not produce
intragap states, as shown in Fig.\,\ref{Dos} (c) .  Instead, surface
N-$p$ states hybridize with the O-$p$ of the functional group. This
results in states directly at the VBM.  The states associated with the
C-$p$ orbitals are located deep in the valence band and overlap with
Ga-$d$ states. Molecular states appear in the conduction band only at
very high energies and stem from C and O $p$-states. Functionalization
with -${\rm NH_{2}}$ groups does not lead to significant changes in
the GaN band structure. The most striking feature is perhaps an
overall shift to higher energies compared to the previous
systems. This might be explained noticing the non-dissociative binding
mode of this molecule. The surface nitrogen atoms are not saturated
and thus possess dangling bonds, which strongly reduces the work
function of the functionalized system.  Our results are in agreement
with experimental findings, which suggested that the formation of
amine groups do not significantly alter the conductivity of the GaN
substrate\cite{Stine}.

The above discussed results on the electronic structure of
functionalized GaN surfaces are very similar to earlier investigations
on ZnO surfaces\,\cite{Prev1,Prev2}. In order to identify further
characteristics and similarities with ZnO, we have calculated the
optical properties of the GaN functionalized surfaces. The dielectric
function for these systems is shown in Fig.~\ref{fig:diel}. While for
functionalized ZnO the intra-gap states of the thiol group are
optically active and lead to an enhancement of the low energetic
absorption\cite{Prev1,Prev2}, no such behavior is found for thiol
groups on GaN surfaces. This means that such a functionalization would
trap holes in the band structure which do not recombine. A similar
behavior is found for -COOH on GaN, where also the presence
of the molecular states at the band-edge suppresses the optical
absorption.  In contrast, surfaces modified with -${\rm NH_{2}}$ and bare surfaces show a distinct peak around the absorption onset, reflecting the presence
of dangling bonds.

\section{Conclusions}

In summary, we have investigated the structural and electronic
properties of GaN surfaces modified by small ligands. We found that
the functionalization with carboxilic acids or amine groups does not
influence the electronic properties of these surfaces significantly.
Functionalization with thiol groups resulted in the appearance of
intra-gap molecular states. However, in contrast to thiol groups on
previously investigated ZnO surfaces\cite{JPCCaccepted}, these states
are not optically active, probably leading to hole trapping. Further
investigations on more realistic surfaces including perhaps the
presence of hydroxyl groups and defects are needed to suggest whether
and how surface modification can enhance the optical properties of
GaN.

\section{acknowledgement}
The authors acknowledge funding by the DFG research group FOR 1616 Dynamics and 
Interactions of Semiconductor Nanowires for Optoelectronics. A. L. Rosa also thanks CNPq and FAPEG for funding.
\section{acknowledgement}

\bibliographystyle{apsrev4-1}

\begin{thebibliography}{31}%
\makeatletter
\providecommand \@ifxundefined [1]{%
 \@ifx{#1\undefined}
}%
\providecommand \@ifnum [1]{%
 \ifnum #1\expandafter \@firstoftwo
 \else \expandafter \@secondoftwo
 \fi
}%
\providecommand \@ifx [1]{%
 \ifx #1\expandafter \@firstoftwo
 \else \expandafter \@secondoftwo
 \fi
}%
\providecommand \natexlab [1]{#1}%
\providecommand \enquote  [1]{``#1''}%
\providecommand \bibnamefont  [1]{#1}%
\providecommand \bibfnamefont [1]{#1}%
\providecommand \citenamefont [1]{#1}%
\providecommand \href@noop [0]{\@secondoftwo}%
\providecommand \href [0]{\begingroup \@sanitize@url \@href}%
\providecommand \@href[1]{\@@startlink{#1}\@@href}%
\providecommand \@@href[1]{\endgroup#1\@@endlink}%
\providecommand \@sanitize@url [0]{\catcode `\\12\catcode `\$12\catcode
  `\&12\catcode `\#12\catcode `\^12\catcode `\_12\catcode `\%12\relax}%
\providecommand \@@startlink[1]{}%
\providecommand \@@endlink[0]{}%
\providecommand \url  [0]{\begingroup\@sanitize@url \@url }%
\providecommand \@url [1]{\endgroup\@href {#1}{\urlprefix }}%
\providecommand \urlprefix  [0]{URL }%
\providecommand \Eprint [0]{\href }%
\providecommand \doibase [0]{http://dx.doi.org/}%
\providecommand \selectlanguage [0]{\@gobble}%
\providecommand \bibinfo  [0]{\@secondoftwo}%
\providecommand \bibfield  [0]{\@secondoftwo}%
\providecommand \translation [1]{[#1]}%
\providecommand \BibitemOpen [0]{}%
\providecommand \bibitemStop [0]{}%
\providecommand \bibitemNoStop [0]{.\EOS\space}%
\providecommand \EOS [0]{\spacefactor3000\relax}%
\providecommand \BibitemShut  [1]{\csname bibitem#1\endcsname}%
\let\auto@bib@innerbib\@empty
\bibitem [{\citenamefont {Chang}\ \emph {et~al.}(2007)\citenamefont {Chang},
  \citenamefont {Chien}, \citenamefont {Stichtenoth}, \citenamefont {Ronning},\
  and\ \citenamefont {Lu}}]{STVR}%
  \BibitemOpen
  \bibfield  {author} {\bibinfo {author} {\bibfnamefont {P.-C.}\ \bibnamefont
  {Chang}}, \bibinfo {author} {\bibfnamefont {C.-J.}\ \bibnamefont {Chien}},
  \bibinfo {author} {\bibfnamefont {D.}~\bibnamefont {Stichtenoth}}, \bibinfo
  {author} {\bibfnamefont {C.}~\bibnamefont {Ronning}}, \ and\ \bibinfo
  {author} {\bibfnamefont {J.~G.}\ \bibnamefont {Lu}},\ }\href@noop {}
  {\bibfield  {journal} {\bibinfo  {journal} {Applied Physics Letters}\
  }\textbf {\bibinfo {volume} {90}},\ \bibinfo {pages} {113101} (\bibinfo
  {year} {2007})}\BibitemShut {NoStop}%
\bibitem [{\citenamefont {Shalish}\ \emph {et~al.}(2004)\citenamefont
  {Shalish}, \citenamefont {Temkin},\ and\ \citenamefont
  {Narayanamurti}}]{STVR2}%
  \BibitemOpen
  \bibfield  {author} {\bibinfo {author} {\bibfnamefont {I.}~\bibnamefont
  {Shalish}}, \bibinfo {author} {\bibfnamefont {H.}~\bibnamefont {Temkin}}, \
  and\ \bibinfo {author} {\bibfnamefont {V.}~\bibnamefont {Narayanamurti}},\
  }\href@noop {} {\bibfield  {journal} {\bibinfo  {journal} {Phys. Rev. B}\
  }\textbf {\bibinfo {volume} {69}},\ \bibinfo {pages} {245401} (\bibinfo
  {year} {2004})}\BibitemShut {NoStop}%
\bibitem [{\citenamefont {Nakamura}\ \emph {et~al.}(1993)\citenamefont
  {Nakamura}, \citenamefont {Senoh},\ and\ \citenamefont {Mukai}}]{Nakamura}%
  \BibitemOpen
  \bibfield  {author} {\bibinfo {author} {\bibfnamefont {S.}~\bibnamefont
  {Nakamura}}, \bibinfo {author} {\bibfnamefont {M.}~\bibnamefont {Senoh}}, \
  and\ \bibinfo {author} {\bibfnamefont {T.}~\bibnamefont {Mukai}},\
  }\href@noop {} {\bibfield  {journal} {\bibinfo  {journal} {Applied Physics
  Letters}\ }\textbf {\bibinfo {volume} {62}},\ \bibinfo {pages} {2390}
  (\bibinfo {year} {1993})}\BibitemShut {NoStop}%
\bibitem [{\citenamefont {Morkoç}\ \emph {et~al.}(1994)\citenamefont
  {Morkoç}, \citenamefont {Strite}, \citenamefont {Gao}, \citenamefont {Lin},
  \citenamefont {Sverdlov},\ and\ \citenamefont {Burns}}]{Morkoc}%
  \BibitemOpen
  \bibfield  {author} {\bibinfo {author} {\bibfnamefont {H.}~\bibnamefont
  {Morkoç}}, \bibinfo {author} {\bibfnamefont {S.}~\bibnamefont {Strite}},
  \bibinfo {author} {\bibfnamefont {G.~B.}\ \bibnamefont {Gao}}, \bibinfo
  {author} {\bibfnamefont {M.~E.}\ \bibnamefont {Lin}}, \bibinfo {author}
  {\bibfnamefont {B.}~\bibnamefont {Sverdlov}}, \ and\ \bibinfo {author}
  {\bibfnamefont {M.}~\bibnamefont {Burns}},\ }\href@noop {} {\bibfield
  {journal} {\bibinfo  {journal} {Journal of Applied Physics}\ }\textbf
  {\bibinfo {volume} {76}},\ \bibinfo {pages} {1363} (\bibinfo {year}
  {1994})}\BibitemShut {NoStop}%
\bibitem [{\citenamefont {Guo}\ \emph {et~al.}(2010)\citenamefont {Guo},
  \citenamefont {Abdulagatov}, \citenamefont {Rourke}, \citenamefont
  {Bertness}, \citenamefont {George}, \citenamefont {Lee},\ and\ \citenamefont
  {Tan}}]{Guo}%
  \BibitemOpen
  \bibfield  {author} {\bibinfo {author} {\bibfnamefont {D.~J.}\ \bibnamefont
  {Guo}}, \bibinfo {author} {\bibfnamefont {A.~I.}\ \bibnamefont
  {Abdulagatov}}, \bibinfo {author} {\bibfnamefont {D.~M.}\ \bibnamefont
  {Rourke}}, \bibinfo {author} {\bibfnamefont {K.~A.}\ \bibnamefont
  {Bertness}}, \bibinfo {author} {\bibfnamefont {S.~M.}\ \bibnamefont
  {George}}, \bibinfo {author} {\bibfnamefont {Y.~C.}\ \bibnamefont {Lee}}, \
  and\ \bibinfo {author} {\bibfnamefont {W.}~\bibnamefont {Tan}},\ }\href@noop
  {} {\bibfield  {journal} {\bibinfo  {journal} {Langmuir}\ }\textbf {\bibinfo
  {volume} {26}},\ \bibinfo {pages} {18382} (\bibinfo {year}
  {2010})}\BibitemShut {NoStop}%
\bibitem [{\citenamefont {Kim}\ \emph {et~al.}(2006)\citenamefont {Kim},
  \citenamefont {Colavita}, \citenamefont {Metz}, \citenamefont {andB. Sun},
  \citenamefont {Uhlrich}, \citenamefont {Wang}, \citenamefont {Kuech},\ and\
  \citenamefont {Hamers}}]{Kim2006}%
  \BibitemOpen
  \bibfield  {author} {\bibinfo {author} {\bibfnamefont {H.}~\bibnamefont
  {Kim}}, \bibinfo {author} {\bibfnamefont {P.~E.}\ \bibnamefont {Colavita}},
  \bibinfo {author} {\bibfnamefont {K.~M.}\ \bibnamefont {Metz}}, \bibinfo
  {author} {\bibfnamefont {B.~M.~N.}\ \bibnamefont {andB. Sun}}, \bibinfo
  {author} {\bibfnamefont {J.}~\bibnamefont {Uhlrich}}, \bibinfo {author}
  {\bibfnamefont {X.~Y.}\ \bibnamefont {Wang}}, \bibinfo {author}
  {\bibfnamefont {T.~F.}\ \bibnamefont {Kuech}}, \ and\ \bibinfo {author}
  {\bibfnamefont {R.~J.}\ \bibnamefont {Hamers}},\ }\href@noop {} {\bibfield
  {journal} {\bibinfo  {journal} {Langmuir}\ }\textbf {\bibinfo {volume}
  {22}},\ \bibinfo {pages} {8121} (\bibinfo {year} {2006})}\BibitemShut
  {NoStop}%
\bibitem [{\citenamefont {Williams}\ \emph {et~al.}(2014)\citenamefont
  {Williams}, \citenamefont {Davydov}, \citenamefont {Oleshko}, \citenamefont
  {Steffens}, \citenamefont {Levin}, \citenamefont {Lin}, \citenamefont
  {Bertness}, \citenamefont {Manocchi}, \citenamefont {Schreifels},\ and\
  \citenamefont {Rao}}]{Williams}%
  \BibitemOpen
  \bibfield  {author} {\bibinfo {author} {\bibfnamefont {E.~H.}\ \bibnamefont
  {Williams}}, \bibinfo {author} {\bibfnamefont {A.~V.}\ \bibnamefont
  {Davydov}}, \bibinfo {author} {\bibfnamefont {V.~P.}\ \bibnamefont
  {Oleshko}}, \bibinfo {author} {\bibfnamefont {K.~L.}\ \bibnamefont
  {Steffens}}, \bibinfo {author} {\bibfnamefont {I.}~\bibnamefont {Levin}},
  \bibinfo {author} {\bibfnamefont {N.~J.}\ \bibnamefont {Lin}}, \bibinfo
  {author} {\bibfnamefont {K.~A.}\ \bibnamefont {Bertness}}, \bibinfo {author}
  {\bibfnamefont {A.~K.}\ \bibnamefont {Manocchi}}, \bibinfo {author}
  {\bibfnamefont {J.~A.}\ \bibnamefont {Schreifels}}, \ and\ \bibinfo {author}
  {\bibfnamefont {M.~V.}\ \bibnamefont {Rao}},\ }\href@noop {} {\bibfield
  {journal} {\bibinfo  {journal} {Surf. Sci}\ }\textbf {\bibinfo {volume}
  {627}},\ \bibinfo {pages} {23} (\bibinfo {year} {2014})}\BibitemShut
  {NoStop}%
\bibitem [{\citenamefont {Choi}\ \emph {et~al.}(2017)\citenamefont {Choi},
  \citenamefont {Kim}, \citenamefont {Jeong}, \citenamefont {Lim},
  \citenamefont {Lee}, \citenamefont {Umar},\ and\ \citenamefont
  {Lee}}]{Choi2017}%
  \BibitemOpen
  \bibfield  {author} {\bibinfo {author} {\bibfnamefont {M.-K.}\ \bibnamefont
  {Choi}}, \bibinfo {author} {\bibfnamefont {G.-S.}\ \bibnamefont {Kim}},
  \bibinfo {author} {\bibfnamefont {J.-T.}\ \bibnamefont {Jeong}}, \bibinfo
  {author} {\bibfnamefont {J.-T.}\ \bibnamefont {Lim}}, \bibinfo {author}
  {\bibfnamefont {W.-Y.}\ \bibnamefont {Lee}}, \bibinfo {author} {\bibfnamefont
  {A.}~\bibnamefont {Umar}}, \ and\ \bibinfo {author} {\bibfnamefont {S.-K.}\
  \bibnamefont {Lee}},\ }\href@noop {} {\bibfield  {journal} {\bibinfo
  {journal} {Sci. Rep.}\ }\textbf {\bibinfo {volume} {7}},\ \bibinfo {pages}
  {14917} (\bibinfo {year} {2017})}\BibitemShut {NoStop}%
\bibitem [{\citenamefont {Bermudez}(2003)}]{Bermudez}%
  \BibitemOpen
  \bibfield  {author} {\bibinfo {author} {\bibfnamefont {V.~M.}\ \bibnamefont
  {Bermudez}},\ }\href@noop {} {\bibfield  {journal} {\bibinfo  {journal}
  {Langmuir}\ ,\ \bibinfo {pages} {6813}} (\bibinfo {year} {2003})}\BibitemShut
  {NoStop}%
\bibitem [{\citenamefont {Schwarz}\ \emph {et~al.}(2013)\citenamefont
  {Schwarz}, \citenamefont {Cimalla}, \citenamefont {Eichapfel}, \citenamefont
  {Himmerlich}, \citenamefont {Krischok},\ and\ \citenamefont
  {Ambacher}}]{Ambacher}%
  \BibitemOpen
  \bibfield  {author} {\bibinfo {author} {\bibfnamefont {S.~U.}\ \bibnamefont
  {Schwarz}}, \bibinfo {author} {\bibfnamefont {V.}~\bibnamefont {Cimalla}},
  \bibinfo {author} {\bibfnamefont {G.}~\bibnamefont {Eichapfel}}, \bibinfo
  {author} {\bibfnamefont {M.}~\bibnamefont {Himmerlich}}, \bibinfo {author}
  {\bibfnamefont {S.}~\bibnamefont {Krischok}}, \ and\ \bibinfo {author}
  {\bibfnamefont {O.}~\bibnamefont {Ambacher}},\ }\href@noop {} {\bibfield
  {journal} {\bibinfo  {journal} {Langmuir}\ }\textbf {\bibinfo {volume}
  {29}},\ \bibinfo {pages} {6296} (\bibinfo {year} {2013})}\BibitemShut
  {NoStop}%
\bibitem [{\citenamefont {Arisio}\ \emph {et~al.}(2013)\citenamefont {Arisio},
  \citenamefont {Cassou},\ and\ \citenamefont {Lieberman}}]{Lieberman}%
  \BibitemOpen
  \bibfield  {author} {\bibinfo {author} {\bibfnamefont {C.}~\bibnamefont
  {Arisio}}, \bibinfo {author} {\bibfnamefont {C.~A.}\ \bibnamefont {Cassou}},
  \ and\ \bibinfo {author} {\bibfnamefont {M.}~\bibnamefont {Lieberman}},\
  }\href@noop {} {\bibfield  {journal} {\bibinfo  {journal} {Langmuir}\
  }\textbf {\bibinfo {volume} {29}},\ \bibinfo {pages} {5145} (\bibinfo {year}
  {2013})}\BibitemShut {NoStop}%
\bibitem [{\citenamefont {Stine}\ \emph {et~al.}(2010)\citenamefont {Stine},
  \citenamefont {Simpkins}, \citenamefont {Mulvaney}, \citenamefont {Whitman},\
  and\ \citenamefont {Tamanaha}}]{Stine}%
  \BibitemOpen
  \bibfield  {author} {\bibinfo {author} {\bibfnamefont {R.}~\bibnamefont
  {Stine}}, \bibinfo {author} {\bibfnamefont {B.}~\bibnamefont {Simpkins}},
  \bibinfo {author} {\bibfnamefont {S.~P.}\ \bibnamefont {Mulvaney}}, \bibinfo
  {author} {\bibfnamefont {L.~J.}\ \bibnamefont {Whitman}}, \ and\ \bibinfo
  {author} {\bibfnamefont {C.~R.}\ \bibnamefont {Tamanaha}},\ }\href@noop {}
  {\bibfield  {journal} {\bibinfo  {journal} {Appl. Surf. Sci.}\ }\textbf
  {\bibinfo {volume} {256}},\ \bibinfo {pages} {4171} (\bibinfo {year}
  {2010})}\BibitemShut {NoStop}%
\bibitem [{\citenamefont {Perdew}\ \emph {et~al.}(1996)\citenamefont {Perdew},
  \citenamefont {Burke},\ and\ \citenamefont {Ernzerhof}}]{Perdew:96}%
  \BibitemOpen
  \bibfield  {author} {\bibinfo {author} {\bibfnamefont {J.~P.}\ \bibnamefont
  {Perdew}}, \bibinfo {author} {\bibfnamefont {K.}~\bibnamefont {Burke}}, \
  and\ \bibinfo {author} {\bibfnamefont {M.}~\bibnamefont {Ernzerhof}},\
  }\href@noop {} {\bibfield  {journal} {\bibinfo  {journal} {Phys. Rev. Lett}\
  }\textbf {\bibinfo {volume} {77}},\ \bibinfo {pages} {3865} (\bibinfo {year}
  {1996})}\BibitemShut {NoStop}%
\bibitem [{\citenamefont {Perdew}\ \emph {et~al.}(1997)\citenamefont {Perdew},
  \citenamefont {Burke},\ and\ \citenamefont {Ernzerhof}}]{Perdew:97}%
  \BibitemOpen
  \bibfield  {author} {\bibinfo {author} {\bibfnamefont {J.~P.}\ \bibnamefont
  {Perdew}}, \bibinfo {author} {\bibfnamefont {K.}~\bibnamefont {Burke}}, \
  and\ \bibinfo {author} {\bibfnamefont {M.}~\bibnamefont {Ernzerhof}},\
  }\href@noop {} {\bibfield  {journal} {\bibinfo  {journal} {Phys. Rev. Lett.}\
  }\textbf {\bibinfo {volume} {78}},\ \bibinfo {pages} {1396} (\bibinfo {year}
  {1997})}\BibitemShut {NoStop}%
\bibitem [{\citenamefont {Gavrilenko}\ and\ \citenamefont {Wu}(2000)}]{GGAGap}%
  \BibitemOpen
  \bibfield  {author} {\bibinfo {author} {\bibfnamefont {V.~I.}\ \bibnamefont
  {Gavrilenko}}\ and\ \bibinfo {author} {\bibfnamefont {R.~Q.}\ \bibnamefont
  {Wu}},\ }\href {\doibase 10.1103/PhysRevB.61.2632} {\bibfield  {journal}
  {\bibinfo  {journal} {Phys. Rev. B}\ }\textbf {\bibinfo {volume} {61}},\
  \bibinfo {pages} {2632} (\bibinfo {year} {2000})}\BibitemShut {NoStop}%
\bibitem [{\citenamefont {Arora}\ \emph {et~al.}(2015)\citenamefont {Arora},
  \citenamefont {Mund}, \citenamefont {Sharma}, \citenamefont {Heda},\ and\
  \citenamefont {Ahuja}}]{GGAGap1}%
  \BibitemOpen
  \bibfield  {author} {\bibinfo {author} {\bibfnamefont {G.}~\bibnamefont
  {Arora}}, \bibinfo {author} {\bibfnamefont {H.~S.}\ \bibnamefont {Mund}},
  \bibinfo {author} {\bibfnamefont {V.}~\bibnamefont {Sharma}}, \bibinfo
  {author} {\bibfnamefont {N.~L.}\ \bibnamefont {Heda}}, \ and\ \bibinfo
  {author} {\bibfnamefont {B.~L.}\ \bibnamefont {Ahuja}},\ }\href@noop {}
  {\bibfield  {journal} {\bibinfo  {journal} {Indian J. Pure Appl. Phys.}\
  }\textbf {\bibinfo {volume} {53}},\ \bibinfo {pages} {328} (\bibinfo {year}
  {2015})}\BibitemShut {NoStop}%
\bibitem [{\citenamefont {Haynes}(2016)}]{CRC}%
  \BibitemOpen
  \bibinfo {editor} {\bibfnamefont {W.~M.}\ \bibnamefont {Haynes}},\ ed.,\
  \enquote {\bibinfo {title} {Crc handbook of chemistry and physics: A
  ready-reference book of chemical and physical data},}\ \ (\bibinfo
  {publisher} {CRC Press},\ \bibinfo {address} {Boca Raton},\ \bibinfo {year}
  {2016})\ \bibinfo {edition} {97th}\ ed.\BibitemShut {Stop}%
\bibitem [{\citenamefont {Paier}\ \emph {et~al.}(2006)\citenamefont {Paier},
  \citenamefont {Marsman}, \citenamefont {Hummer}, \citenamefont {Kresse},
  \citenamefont {Gerber},\ and\ \citenamefont {Ángyán}}]{Paier}%
  \BibitemOpen
  \bibfield  {author} {\bibinfo {author} {\bibfnamefont {J.}~\bibnamefont
  {Paier}}, \bibinfo {author} {\bibfnamefont {M.}~\bibnamefont {Marsman}},
  \bibinfo {author} {\bibfnamefont {K.}~\bibnamefont {Hummer}}, \bibinfo
  {author} {\bibfnamefont {G.}~\bibnamefont {Kresse}}, \bibinfo {author}
  {\bibfnamefont {I.~C.}\ \bibnamefont {Gerber}}, \ and\ \bibinfo {author}
  {\bibfnamefont {J.~G.}\ \bibnamefont {Ángyán}},\ }\href@noop {} {\bibfield
  {journal} {\bibinfo  {journal} {The Journal of Chemical Physics}\ }\textbf
  {\bibinfo {volume} {124}},\ \bibinfo {pages} {154709} (\bibinfo {year}
  {2006})}\BibitemShut {NoStop}%
\bibitem [{\citenamefont {Adamo}\ and\ \citenamefont {Barone}(1999)}]{PBE0}%
  \BibitemOpen
  \bibfield  {author} {\bibinfo {author} {\bibfnamefont {C.}~\bibnamefont
  {Adamo}}\ and\ \bibinfo {author} {\bibfnamefont {V.}~\bibnamefont {Barone}},\
  }\href@noop {} {\bibfield  {journal} {\bibinfo  {journal} {The Journal of
  Chemical Physics}\ }\textbf {\bibinfo {volume} {110}},\ \bibinfo {pages}
  {6158} (\bibinfo {year} {1999})}\BibitemShut {NoStop}%
\bibitem [{\citenamefont {Hohenberg}\ and\ \citenamefont
  {Kohn}(1964)}]{Hohenberg1964}%
  \BibitemOpen
  \bibfield  {author} {\bibinfo {author} {\bibfnamefont {P.}~\bibnamefont
  {Hohenberg}}\ and\ \bibinfo {author} {\bibfnamefont {W.}~\bibnamefont
  {Kohn}},\ }\href {\doibase 10.1103/PhysRev.136.B864} {\bibfield  {journal}
  {\bibinfo  {journal} {Phys. Rev.}\ }\textbf {\bibinfo {volume} {136}},\
  \bibinfo {pages} {B864} (\bibinfo {year} {1964})}\BibitemShut {NoStop}%
\bibitem [{\citenamefont {Kohn}\ and\ \citenamefont {Sham}(1965)}]{Kohn1965}%
  \BibitemOpen
  \bibfield  {author} {\bibinfo {author} {\bibfnamefont {W.}~\bibnamefont
  {Kohn}}\ and\ \bibinfo {author} {\bibfnamefont {L.~J.}\ \bibnamefont
  {Sham}},\ }\href {\doibase 10.1103/PhysRev.140.A1133} {\bibfield  {journal}
  {\bibinfo  {journal} {Phys. Rev.}\ }\textbf {\bibinfo {volume} {140}},\
  \bibinfo {pages} {A1133} (\bibinfo {year} {1965})}\BibitemShut {NoStop}%
\bibitem [{\citenamefont {Kresse}\ and\ \citenamefont
  {Furthm\"uller}(1996{\natexlab{a}})}]{VASP:3}%
  \BibitemOpen
  \bibfield  {author} {\bibinfo {author} {\bibfnamefont {G.}~\bibnamefont
  {Kresse}}\ and\ \bibinfo {author} {\bibfnamefont {J.}~\bibnamefont
  {Furthm\"uller}},\ }\href@noop {} {\bibfield  {journal} {\bibinfo  {journal}
  {Comput. Mat. Sci.}\ }\textbf {\bibinfo {volume} {6}},\ \bibinfo {pages} {15}
  (\bibinfo {year} {1996}{\natexlab{a}})}\BibitemShut {NoStop}%
\bibitem [{\citenamefont {Kresse}\ and\ \citenamefont
  {Furthm\"uller}(1996{\natexlab{b}})}]{VASP:4}%
  \BibitemOpen
  \bibfield  {author} {\bibinfo {author} {\bibfnamefont {G.}~\bibnamefont
  {Kresse}}\ and\ \bibinfo {author} {\bibfnamefont {J.}~\bibnamefont
  {Furthm\"uller}},\ }\href@noop {} {\bibfield  {journal} {\bibinfo  {journal}
  {Phys. Rev. B}\ }\textbf {\bibinfo {volume} {54}},\ \bibinfo {pages} {11169}
  (\bibinfo {year} {1996}{\natexlab{b}})}\BibitemShut {NoStop}%
\bibitem [{\citenamefont {Kresse}\ and\ \citenamefont
  {Joubert}(1999)}]{Kresse:99}%
  \BibitemOpen
  \bibfield  {author} {\bibinfo {author} {\bibfnamefont {G.}~\bibnamefont
  {Kresse}}\ and\ \bibinfo {author} {\bibfnamefont {D.}~\bibnamefont
  {Joubert}},\ }\href@noop {} {\bibfield  {journal} {\bibinfo  {journal} {Phys.
  Rev. B}\ }\textbf {\bibinfo {volume} {59}},\ \bibinfo {pages} {1758}
  (\bibinfo {year} {1999})}\BibitemShut {NoStop}%
\bibitem [{\citenamefont {Bl\"ochl}(1994)}]{Bloechel:94}%
  \BibitemOpen
  \bibfield  {author} {\bibinfo {author} {\bibfnamefont {P.~E.}\ \bibnamefont
  {Bl\"ochl}},\ }\href {\doibase 10.1103/PhysRevB.50.17953} {\bibfield
  {journal} {\bibinfo  {journal} {Phys. Rev. B}\ }\textbf {\bibinfo {volume}
  {50}},\ \bibinfo {pages} {17953} (\bibinfo {year} {1994})}\BibitemShut
  {NoStop}%
\bibitem [{\citenamefont {Dominguez}\ \emph {et~al.}(2014)\citenamefont
  {Dominguez}, \citenamefont {Lorke}, \citenamefont {Schoenhalz}, \citenamefont
  {Rosa}, \citenamefont {Frauenheim}, \citenamefont {Rocha},\ and\
  \citenamefont {Dalpian}}]{Prev1}%
  \BibitemOpen
  \bibfield  {author} {\bibinfo {author} {\bibfnamefont {A.}~\bibnamefont
  {Dominguez}}, \bibinfo {author} {\bibfnamefont {M.}~\bibnamefont {Lorke}},
  \bibinfo {author} {\bibfnamefont {A.~L.}\ \bibnamefont {Schoenhalz}},
  \bibinfo {author} {\bibfnamefont {A.~L.}\ \bibnamefont {Rosa}}, \bibinfo
  {author} {\bibfnamefont {T.}~\bibnamefont {Frauenheim}}, \bibinfo {author}
  {\bibfnamefont {A.~R.}\ \bibnamefont {Rocha}}, \ and\ \bibinfo {author}
  {\bibfnamefont {G.~M.}\ \bibnamefont {Dalpian}},\ }\href@noop {} {\bibfield
  {journal} {\bibinfo  {journal} {Journal of Applied Physics}\ }\textbf
  {\bibinfo {volume} {115}},\ \bibinfo {pages} {203720} (\bibinfo {year}
  {2014})}\BibitemShut {NoStop}%
\bibitem [{\citenamefont {Moreira}\ \emph {et~al.}(2012)\citenamefont
  {Moreira}, \citenamefont {Dominguez}, \citenamefont {Frauenheim},\ and\
  \citenamefont {da~Rosa}}]{Prev2}%
  \BibitemOpen
  \bibfield  {author} {\bibinfo {author} {\bibfnamefont {N.~H.}\ \bibnamefont
  {Moreira}}, \bibinfo {author} {\bibfnamefont {A.}~\bibnamefont {Dominguez}},
  \bibinfo {author} {\bibfnamefont {T.}~\bibnamefont {Frauenheim}}, \ and\
  \bibinfo {author} {\bibfnamefont {A.~L.}\ \bibnamefont {da~Rosa}},\
  }\href@noop {} {\bibfield  {journal} {\bibinfo  {journal} {Phys. Chem. Chem.
  Phys.}\ }\textbf {\bibinfo {volume} {14}},\ \bibinfo {pages} {15445}
  (\bibinfo {year} {2012})}\BibitemShut {NoStop}%
\bibitem [{\citenamefont {Umari}\ \emph {et~al.}(2013)\citenamefont {Umari},
  \citenamefont {Giacomazzi}, \citenamefont {De~Angelis}, \citenamefont
  {Pastore},\ and\ \citenamefont {Baroni}}]{Align1}%
  \BibitemOpen
  \bibfield  {author} {\bibinfo {author} {\bibfnamefont {P.}~\bibnamefont
  {Umari}}, \bibinfo {author} {\bibfnamefont {L.}~\bibnamefont {Giacomazzi}},
  \bibinfo {author} {\bibfnamefont {F.}~\bibnamefont {De~Angelis}}, \bibinfo
  {author} {\bibfnamefont {M.}~\bibnamefont {Pastore}}, \ and\ \bibinfo
  {author} {\bibfnamefont {S.}~\bibnamefont {Baroni}},\ }\href@noop {}
  {\bibfield  {journal} {\bibinfo  {journal} {The Journal of Chemical Physics}\
  }\textbf {\bibinfo {volume} {139}},\ \bibinfo {pages} {014709} (\bibinfo
  {year} {2013})}\BibitemShut {NoStop}%
\bibitem [{\citenamefont {Kang}\ \emph {et~al.}(2013)\citenamefont {Kang},
  \citenamefont {Tongay}, \citenamefont {Zhou}, \citenamefont {Li},\ and\
  \citenamefont {Wu}}]{Align2}%
  \BibitemOpen
  \bibfield  {author} {\bibinfo {author} {\bibfnamefont {J.}~\bibnamefont
  {Kang}}, \bibinfo {author} {\bibfnamefont {S.}~\bibnamefont {Tongay}},
  \bibinfo {author} {\bibfnamefont {J.}~\bibnamefont {Zhou}}, \bibinfo {author}
  {\bibfnamefont {J.}~\bibnamefont {Li}}, \ and\ \bibinfo {author}
  {\bibfnamefont {J.}~\bibnamefont {Wu}},\ }\href@noop {} {\bibfield  {journal}
  {\bibinfo  {journal} {Applied Physics Letters}\ }\textbf {\bibinfo {volume}
  {102}},\ \bibinfo {pages} {012111} (\bibinfo {year} {2013})}\BibitemShut
  {NoStop}%
\bibitem [{\citenamefont {Garza}\ and\ \citenamefont
  {Scuseria}(2016)}]{Scuseria}%
  \BibitemOpen
  \bibfield  {author} {\bibinfo {author} {\bibfnamefont {A.~J.}\ \bibnamefont
  {Garza}}\ and\ \bibinfo {author} {\bibfnamefont {G.~E.}\ \bibnamefont
  {Scuseria}},\ }\href@noop {} {\bibfield  {journal} {\bibinfo  {journal} {J.
  Phys. Chem. Lett.}\ }\textbf {\bibinfo {volume} {7}},\ \bibinfo {pages}
  {4165} (\bibinfo {year} {2016})}\BibitemShut {NoStop}%
\bibitem [{\citenamefont {Franke}\ \emph {et~al.}(2018)\citenamefont {Franke},
  \citenamefont {Lorke}, \citenamefont {Rosa},\ and\ \citenamefont
  {Frauenheim}}]{JPCCaccepted}%
  \BibitemOpen
  \bibfield  {author} {\bibinfo {author} {\bibfnamefont {D.}~\bibnamefont
  {Franke}}, \bibinfo {author} {\bibfnamefont {M.}~\bibnamefont {Lorke}},
  \bibinfo {author} {\bibfnamefont {A.~L.}\ \bibnamefont {Rosa}}, \ and\
  \bibinfo {author} {\bibfnamefont {T.}~\bibnamefont {Frauenheim}},\
  }\href@noop {} {\bibfield  {journal} {\bibinfo  {journal} {J. Phys. Chem. C}\
  } (\bibinfo {year} {2018})},\ \bibinfo {note} {accepted}\BibitemShut
  {NoStop}%
\end{thebibliography}

%
\end{document}